\documentclass[%
 reprint,
 amsmath,amssymb,
 aps,
]{revtex4-2}

\usepackage{graphicx}
\usepackage{dcolumn}
\usepackage{bm}
\usepackage{xcolor}
\usepackage{float}
\usepackage[colorlinks=true, citecolor=blue,linkcolor=red, urlcolor=blue]{hyperref}

\begin{document}

\title{Improved Convergence of Carleman-Embedded Quantum Algorithm for the Vlasov–Poisson System}

\author{Matthew Christensen}
\affiliation{Department of Physics, University of Warwick, Coventry CV4 7AL, UK.}

\author{Tom Goffrey}
\affiliation{Department of Physics, University of Warwick, Coventry CV4 7AL, UK.}

\author{Animesh Datta}
\affiliation{Department of Physics, University of Warwick, Coventry CV4 7AL, UK.}

\date{\today}

\begin{abstract}

We extend the regime of convergence of Carleman-embedded quantum algorithms that solve the Vlasov-Poisson equations from kinetic plasma physics.
We establish convergence, using both analytical and numerical lower bounds, for physically reasonable collision frequencies using a Fourier-Hermite expansion of the shifted phase-space distribution function.
We also show that for a large class of basis functions, the convergence of the Carleman-embedded Vlasov-Poisson system requires increasing collision frequency strength with velocity resolution. The complexity of the quantum algorithm depends strongly on whether we seek time-averaged or -resolved outputs.
\end{abstract}

\maketitle


\raggedbottom

\textit{Introduction:}
Plasma and fusion physics is one of the most computationally intensive disciplines.
This is due to the multiscale nature of the phenomena whereby simulations on the longer fluid time and length scales require closure models which can only be provided by the much shorter kinetic scale plasma phenomena \cite{atzeni2005fluid}. Typically, fluid simulations replace the kinetic physics with approximate models, such as the use of flux-limited thermal conduction~\cite{arber2023models} or non-local models~\cite{schurtz2000nonlocal}. However, these models include free parameters, potentially damaging the predictive capability of the simulations and require tuning to accurately reproduce fully kinetic results~\cite{brodrick2017testing}. A truly predictive fluid simulation would require inline kinetic calculations to provide self-consistent evolution of transport coefficients.
These kinetic phenomena are governed by a coupled set of nonlinear partial differential equations known as the Vlasov-Maxwell system~\cite{chen2015introduction}. However, solving them in 3+3 phase-space dimensions is not yet viable on classical computational platforms without (often) making significant approximations~\cite{bell2024fast}. 

Quantum computers may offer a new route for solving classically hard computational problems, sometimes gaining exponential improvements when compared to classical counterparts. This speedup can be seen in problems that are quantum by nature, such as quantum chemistry \cite{feynman_simulating_1982} and material science \cite{PhysRevX.8.011044}, but also problems such as integer factorization \cite{shor1999polynomial} and solving a system of linear equations~\cite{harrow2009quantum} that are not inherently quantum tasks.

Quantum algorithms solving for plasma kinetics in the electrostatic limit governed by the Vlasov-Poisson system of equations have focused on linearizing the system around a Maxwellian distribution function and only considering the distribution function's perturbations. Engel \emph{et al.} \cite{engel_quantum_2019} used a Fourier expansion in space and a grid-based method in velocity to produce a linear Hamiltonian system that they solve using Hamiltonian simulation techniques \cite{low_optimal_2017}. Ameri \emph{et al.} \cite{ameri_quantum_2023} found that using a Hermite expansion in velocity space created exponentially smaller linear systems for the same error.  However, linearized perturbation dynamics do not faithfully capture the nonlinear physics inherent in several applications~\cite{o1965collisionless, bernstein1957exact}.

The first quantum algorithm for solving the nonlinear Vlasov-Poisson system used a finite difference scheme to discretize phase-space
followed by a method called Carleman embedding~\cite{vaszary2024solving}. This method~\cite{liu2021efficient, forets2017explicit, krovi_improved_2023,jennings2025quantum, costa_further_2025} forms an infinite-dimensional linear system of ODEs out of a finite number of higher order monomials of the nonlinear variables. This linear system can then be truncated and solved using a quantum linear ODE solver \cite{childs_hamiltonian_2012, gilyen2019quantum, an2023linear, harrow2009quantum}. The central task is of establishing a bound on the error due to this truncation, a property known as convergence. This relies on the strength of dissipation in the original system being comparable to its nonlinearity. Vaszary \emph{et al.}~\cite{vaszary2024solving} found that convergence for the Carleman-embedded nonlinear Vlasov-Poisson system could only be guaranteed for an unphysically large dissipation or collision frequency of the plasma.

In this paper, we show that convergence of the Carleman-embedded nonlinear Vlasov-Poisson system governing the evolution of the electron distribution function can be guaranteed for physically relevant collision frequencies. We rely on solving for a distribution function shifted by the Maxwellian equilibrium and numerically optimizing over a recently introduced~\cite{jennings2025quantum} Lyapunov function of the linear part of the shifted system. Our numerical results are obtained using a Fourier-Hermite basis expansion of the shifted distribution function. En route, we reframe the Carleman convergence criteria in terms of the plasma's collision frequency. We also show that, for a large class of basis functions, the convergence of the Carleman-embedded nonlinear Vlasov-Poisson system requires the strength of the collision frequency to increase with resolution in the velocity space. 

We show that the efficiency of our quantum algorithm depends whether we seek time-averaged or resolved Fourier-Hermite coefficients. These coefficients, the velocity moments of the distribution function, are the transport coefficients for the fluid models and motivated our choice of the Fourier-Hermite expansion. That we only need to extract a few entries of the quantum encoded solution is an appealing prospect.

\textit{Collisional Vlasov-Poisson System}:
The 1+1 dimensional Vlasov-Poisson system of equations for the phase-space distribution function $f(x,v,t)$ of the electrons \cite{thomas2012review} with a BGK collision operator \cite{bhatnagar1954model} is given by
\begin{align}\label{vlasov}
    \frac{\partial f}{\partial t} + v \frac{\partial f}{\partial x} - E \frac{\partial f}{\partial v} = -\nu(f- f_{\text{M}}),\\\label{poisson}
    -\frac{\partial E}{\partial x} = \int^{\infty}_{-\infty} f \,dv\ - 1,
\end{align} 
in dimensionless units. The details of these units are in Appendix \ref{plasma physics appendix}.
Here, $x$ is the position, $v$ the velocity, $t$ the time, and $E(x,t)$ the electric field.
The Maxwellian is denoted by $f_{\text{M}}(v)=e^{-v^2/2}/\sqrt{2\pi}$.
The collision frequency $\nu$ captures the dissipation in this system. Physically, 
\begin{equation}\label{nu ub}
    \nu \lesssim  1,
\end{equation}
as argued in Appendix \ref{plasma physics appendix}.

Our objective is to solve for velocity moments of $f(x,v,t)$ in a domain $x \in [0,L]$ or some combinations thereof for $t \in [0,T]$ or averaged over $[0,T].$
In position space, we impose periodic boundary conditions $f(0,v,t)=f(L,v,t).$ In velocity space, we impose $\lim_{|v|\rightarrow\infty}f(x,v,t)=0$.

\textit{Shifted Vlasov-Poisson System}:
We define a new distribution function $\hat{f}\equiv f - f_{\text{M}}.$ 
The time derivative and spatial advection operator remain unaffected by this shift in velocity space, whilst the collision operator becomes $-\nu\hat{f}.$
The linear parts gain a term from the derivative of the Maxwellian function. The shifted equations are then
\begin{align}\label{shifted vlasov}
    \frac{\partial \hat{f}}{\partial t} + v\frac{\partial \hat{f}}{\partial x} &- E\left(\frac{\partial \hat{f}}{\partial v} + vf_{\text{M}} \right) = \nu\hat{f}\\ \label{shifted poisson}
    &-\frac{\partial E}{\partial t} = \int_{-\infty}^{\infty}\hat{f}dv
\end{align}
where the initial condition is now $\hat{f}_{\text{in}}(x,v,0)=f(x,v,0)-f_{\text{M}}(v)$.

\textit{Fourier-Hermite Expansion}:
We use a Fourier-Hermite expansion of the distribution function,
\begin{equation}
    \hat{f}(x,v,t) = \sum_{n,m}C_{n,m}(t)e^{inx}\frac{He_{m}(v)f_\text{M}(v)}{\sqrt{m!}},
\end{equation}
where $He_m(v)=(-1)^me^{v^2/2}\frac{d^m}{dv^m} e^{-v^2/2}$ are the Hermite polynomials.
 Substituting this into Eqs. \eqref{shifted vlasov} \& \eqref{shifted poisson} and using the orthogonality relationships of the Fourier and Hermite functions yields the the system of nonlinear ordinary differential equations (ODEs) for the coefficients 
\begin{equation}\label{shift. VP mode eq}
\begin{split}
    &\dot{C}_{p,r}+ip(\sqrt{r}C_{p,r-1}+\sqrt{r+1}C_{p,r+1})\\&+\frac{i}{p}C_{p,0}\delta_{r,1}+\sum_{s\neq0}\frac{i\sqrt{r}}{s}C_{s,0}C_{p-s,r-1} = -\nu C_{p,r},
\end{split}
\end{equation}
as shown in Appendix \ref{fh derivation}. The summation spans symmetrically over the spatial Fourier modes, with $p= -N_x/2, \cdots,N_x/2$, where we have assumed that $N_x$ to be even. In the velocity space, the Hermite modes range from $r=0,\cdots, N_v$. 
The dimension of the system 
in Eq.~\eqref{shift. VP mode eq}
is $N=(N_x+1)(N_v+1)$. 
For simplicity, we set $L=2\pi$ such that the Fourier wavenumbers are $k_p=p$.

Eq.~\eqref{shift. VP mode eq} can be expressed as
\begin{equation}\label{ode}
    \dot{u}(t)= F_1 u(t) + F_2u(t)^{\otimes2}, \quad u(t)\in\mathbb{C}^{N}
\end{equation}
where $u(t) $ is the vector of $N$ coefficients $C_{p,r}(t)$ in Eq. \eqref{shift. VP mode eq} and initial condition $u_0=u(0)$ is the vector with entries $C_{p,r}(0)$. The matrices $F_1\in \mathbb{C}^{N\times N}$ and $F_2\in \mathbb{C}^{N\times N^2}$ contain the linear (including the collision frequency $\nu$) and nonlinear contributions respectively. 
They are given explicitly in Appendix \ref{mappings}.
$\nu$ controls the strength of dissipation in the nonlinear ODEs in Eq.~\eqref{shift. VP mode eq}, in addition to being a physical parameter subject to \eqref{nu ub}.

\textit{Carleman Embedding}: We use Carleman embedding~\cite{liu2021efficient} to solve the system of nonlinear ODEs in Eq.~\eqref{ode}. An infinite dimensional linear system is formed out of higher order tensor powers of the variables of the nonlinear system as unknowns. The infinite dimensional linear system is then truncated, leaving a finite-dimensional linear system approximating the nonlinear dynamics.
The embedding is said to converge if the error from this truncation decreases exponentially with the level of the Carleman truncation $N_c$ -- the number of higher order tensor products kept in the linear system.

Given $F_1$, if there exists a $N\times N$-dimensional Hermitian positive definite matrix  $P$ satisfying the Lyapunov inequality 
\begin{equation}
\label{lyap}
PF_1+F_1^\dagger P\prec0,
\end{equation}
then a  \textit{sufficient} condition for convergence is~\cite{jennings2025quantum}

\begin{equation}\label{Carleman R}
    R_P = \frac{\|F_2\|_P\|u_0\|_P}{-\mu_P(F_1)}< 1
\end{equation}
Here, the generalized matrix, vector, and log-norms are
\begin{align}
    &\|F_2\|_P=\|P^{1/2}F_2(P^{-1/2}\otimes P^{-1/2})\|_2,\\
    &\|u_0\|_P = \|P^{1/2}u_0\|_2,\\
    &\mu_P = \max_{x\neq0}\Re\left\{\frac{\langle x,F_1x\rangle_P}{\langle x,x\rangle_P}\right\}
\end{align}
where the inner product is $\langle x,y\rangle_P=x^\dagger Py$. The convergence is thus a competition  between the nonlinearity in $F_2$ and the collision frequency in $F_1$.

We rearrange the convergence criteria of \eqref{Carleman R} to obtain a lower bound on the plasma collision frequency
\begin{equation} \label{shifted lb}
    \nu > \mu_P(-i(S+Y)) + \|F_2\|_P \|u_0\|_P \equiv \nu_P.
\end{equation}
where $S$ is the matrix representing the spatial linear advection and $Y$ is the matrix containing contributions from the Maxwellian distribution. See Appendix \ref{mappings} for explicit definitions. 
Any $\nu$ satisfying inequality \eqref{shifted lb} implies $\nu >\mu_P(-i(S+Y)) +\|F_2\|_P\|u_0\|_P \geq \mu_P(-i(S+Y))$, so the log-norm of $F_1$ is,
\begin{equation}
    \mu_P(F_1)=-\nu + \mu_P(-i(S+Y))<0.
\end{equation} Hence, inequality \eqref{lyap} is satisfied for any $P$.

\begin{figure*}
    \centering
    \includegraphics[width=\linewidth]{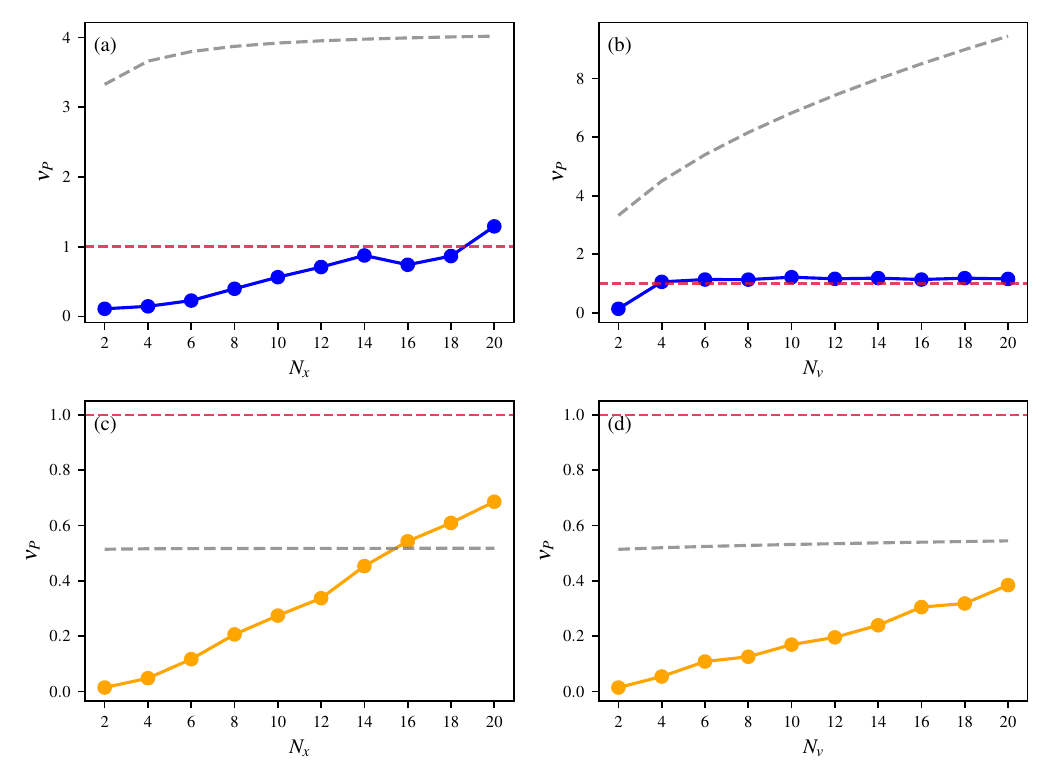}
    \caption{Numerically obtained lower bounds of collision frequency (blue \& yellow), the fixed $P=I$ bounds (grey, dashed) and physical upper limit $\nu =1$ (red, dashed) for a perturbation strength $\alpha =0.01$. (a) Two stream instability initial condition $f_{\text{ts}}$ with fixed $N_v=2$. (b) Two stream instability initial condition with fixed $N_x=2$. (c) Perturbed Maxwellian initial condition $f_{\text{pert}}$ with fixed $N_v=2$. (d) Perturbed Maxwellian initial condition with fixed $N_x=2$. Analytical expressions for the grey dashed lines are given at the end of Appendix \ref{norm evaluations}. Appendix \ref{direct} shows the results of the same minimization for the unshifted Vlasov-Poisson system of Eqs. \eqref{vlasov} \& \eqref{poisson}.}
    \label{fig:shifted comparison}
\end{figure*}

\textit{Convergence of the Carleman Embedding}:
We numerically minimize $\nu_P$ from inequality Eq.~\eqref{shifted lb} over $P$ subject to
inequality \eqref{lyap} for two canonical plasma physics problems: the two stream instability \cite{anderson2001tutorial} and Landau damping \cite{arber2002critical}. 
Their respective initial conditions are denoted $f_{\text{ts}}=(1+\alpha\cos(kx))v^2f_\text{M}$ and $f_{\text{pert}}=(1+\alpha\cos(kx))f_\text{M},$ for a perturbation strength $\alpha$.
Figure \ref{fig:shifted comparison} shows the results. 
We use a Riemannian multi-start method detailed in Appendix \ref{computation details} for the minimization.

The initial condition for the two stream instability describes two streams of electrons that are spatially uniform but travel in opposite directions with average speed of $v=\sqrt{2}$. The blue curves in Figure \ref{fig:shifted comparison} (a), (b) show the Carleman embedding method to converge for smaller values of the collision frequency
compared to the $P=I$ case (grey dashed lines). 
This illustrates the nontrivial effect of minimizing over quadratic Lyapunov functions~\cite{jennings2025quantum} in solving the nonlinear Vlasov-Poisson system.
In Figure \ref{fig:shifted comparison} (a), the minimized bounds eventually surpasses the  physical upper limit in \eqref{nu ub}.

In the more physically relevant case in Figure \ref{fig:shifted comparison} (b), our bound sits slightly above $\nu_P=1$ for increasing $N_v.$ 
Convergence may actually occur for this value given the sufficient nature of the criterion in \eqref{Carleman R}, as has been empirically established for other problems~\cite{liu2021efficient}.

$f_{\text{pert}}$ describes particles that are spatially perturbed by a cosine wave whilst being distributed as Maxwellian in velocity space. This initial condition leads to Landau damping~\cite{case1959plasma}, where the interactions of particles and electrostatic waves cause damping of the electric field. The numerically minimized bounds for this initial condition are shown by the yellow curves in Figure~\ref{fig:shifted comparison} (c), (d). 
Both in the minimized and fixed $P=I$ instances, the bounds fit below the physical limit in~\eqref{nu ub}. We can therefore guarantee convergence of the Carleman embedding method for systems demonstrating Landau damping with a realistic collision frequency.
In Figure~\ref{fig:shifted comparison} (c), for Fourier mode numbers $N_x\geq14$, the minimization fails to improve on the identity Lyapunov matrix cost. This could likely be remedied by increasing the number of initial starts used, or improving the initial point sampling method.

\textit{Identity Lyapunov Matrix}:
The simplest convergence criterion is found for $P=I$ (grey dashed lines in Fig. \ref{fig:shifted comparison}). In this case, the convergence analysis collapses to the earlier results~\cite{krovi_improved_2023}, where the $P$ norms are reduced to spectral, $\ell^2$ and log-norms. 

The lower bound for the shifted problem then becomes
\begin{equation} \label{shifted_converg_identity}
    \nu > \mu(-i(S+Y)) + \|F_2\|_2 \|u_0\|_2 \equiv \nu_I,
\end{equation}
which is evaluated analytically in Appendix \ref{norm evaluations}.
We estimate the number of velocity modes in the expansion before this lower bound exceeds~\eqref{nu ub}. Letting the spatial summation take its upper bound in $\|F_2\|_2$, the bound $\nu_I$ is expressed as 
\begin{equation}
    1>\nu_I\geq \frac{1}{2} + \alpha\sqrt{N_v}\frac{\pi}{\sqrt{6}},
\end{equation}
which we rearrange to get an upper bound
\begin{equation}
    N_v< \frac{3}{2}\pi^{-2}\alpha^{-2}.
\end{equation}
This is the maximum number of velocity modes kept in the expansion before the physical limit in~\eqref{nu ub} is breached. Letting the perturbation strength take the value $\alpha=0.01$, typical of a value used to demonstrate Landau damping \cite{arber2002critical}, the maximum number of velocity modes is $N_v \approx10^{3}$. 
For comparison, Landau damping in 1+3 dimensions has been demonstrated with with $N_x=16$ Fourier modes and $N_v=10$ Hermite modes in each of the velocity dimensions~\cite{delzanno2015multi}. Thus our quantum algorithm for \emph{nonlinear} Landau damping can support a very large number of Hermite modes and an arbitrarily number of Fourier modes before breaching~\eqref{nu ub}.

\textit{Complexity}:
Having shown convergence of the  Carleman embedded approach to solving the shift Vlasov-Poisson system, we now provide a complexity analysis of the quantum algorithm to solve Eq. \eqref{shift. VP mode eq} for $P=I$.

The cost of obtaining a quantum state $\epsilon$-close to
\begin{equation}\label{eq:hist}
    |\psi_T\rangle \propto \sum_{m=1}^{M = T/\Delta t} \|u(m\Delta t)\|_2 |u(m\Delta t)\rangle|m\rangle
\end{equation}
at a time $T$ is~\cite[Theorem 6.4 ]{jennings2025quantum}

\begin{equation} \label{jennings query}
    \mathcal{O}\left( (\alpha_{F_1} + \alpha_{F_2}) \log(1/\epsilon) \log^2(T/\epsilon) \right)
\end{equation}
calls to the state preparation 
and block encodings of $F_1$ and $F_2$. Here, $\alpha_{F_i}=\sqrt{s^cs^r}\|F_i\|_{\max}$ are the rescaling factors required for the block encodings of the matrices $F_i$, where $s^r$ and $s^c$ are the row and column sparisty of the matrices respectively. Since $F_1$ is a tridiagonal matrix with a maximum entry $\|F_1\|_{\max}= \mathcal{O}(N_x\sqrt{N_v})$, its rescaling factor is $\alpha_{F_1}=\mathcal{O}(N_x\sqrt{N_v})$. For the matrix $F_2$, the row sparsity grows as $s^r=N_x$ but has a fixed column sparsity of $s^c=1$. The maximum entry of $\|F_2\|_{\max}=\mathcal{O}(\sqrt{N_v})$, hence its rescaling is $\alpha_{F_2}=\mathcal{O}(\sqrt{N_xN_v})$.

For both of the sparse initial conditions $f_{\text{ts}}$ and $f_{\text{pert}}$ we have used above, the state preparation scales as $\mathcal{O}(\log(N_xN_v))$ \cite{gleinig2021efficient}. To block encode the matrix $F_1$, we require $\mathcal{O}(1)$ queries to the sparse matrix oracles $O^{F_1}_{r}, O^{F_1}_c$ and $O^{F_1}_{A}$ and $\mathcal{O}(\text{polylog}(N_x\sqrt{N_v}/\varepsilon))$ one and two qubit gates. Similarly for the $F_2$ matrix, the block encoding requires $\mathcal{O}(1)$ queries to the $O^{F_2}_{r}, O^{F_2}_c$ and $O^{F_2}_{A}$ sparse matrix oracles and $\mathcal{O}(\text{polylog}(\sqrt{N_xN_v}/\varepsilon))$ additional gates.

Assuming we have access to the oracles, the total query complexity of quantum algorithm producing $|\psi_T\rangle$ in Eq.~\eqref{eq:hist} is
\begin{equation}
\begin{split}
    &\mathcal{O}\left(N_x\sqrt{N_v} \text{polylog}\left(\frac{1}{\varepsilon}, \frac{T}{\varepsilon}, \frac{N_x\sqrt{N_v}}{\varepsilon}\right)\right) \\&= \Tilde{\mathcal{O}}\left(N_x\sqrt{N_v} \right)
\end{split}
\end{equation}
queries to the oracles of $F_1$, $F_2$ and the state preparation of $|u_0\rangle$ ($\tilde{\mathcal{O}}$ omits the polylogarithmic terms). Here, the state preparation produces a state,
\begin{equation}
    |u_0\rangle = \frac{1}{\|u_0\|_2}\sum_{i} [u_0]_i|i\rangle.
\end{equation}

The classical complexity of solving the system in Eq.~\eqref{shift. VP mode eq} is
the number of time steps multiplied by the number of operations per timestep. The number of timesteps $M$ must satisfy the Courant–Friedrichs–Lewy (CFL) condition such that $M=\mathcal{O}(T\|F_1\|_2/\epsilon)=\mathcal{O}(TN_x\sqrt{N_v}/\epsilon)$. For the operations per timestep, we can use an FFT to solve the convolution term captured by $F_2$ with $\mathcal{O}(N_xN_v\log(N_x))$ operations. Therefore, the number of classical operations scale as $\mathcal{O}(TN_x^2N_v^{3/2}\log(N_x)/\epsilon)$.

Temporal averages of physical quantities, such as the heat flux $Q(x,t) = \sum_n 2 (\sqrt{6}C_{n,3}(t) + 3C_{n,1}(t))\cos{nx} ,$
can be efficiently extracted from the history states in Eq.~\eqref{eq:hist}. Thus the quantum algorithm requires
\begin{equation}
    \tilde{\mathcal{O}}\left(\frac{\epsilon}{TN_x\sqrt{N_v}}\right)
\end{equation}
fewer operations than the classical counterpart.

Extracting time-resolved quantities, say at a final timestep $M,$ is more desirable.
Then, idling steps \cite{Berry2014} can be used to increase the probability of successful measurement of the $m=M$ subspace. However, due to the stability requirements of the Carleman embedding and shifting by the Maxwellian, the norm of the solution  at a final timestep $M$ becomes exponentially small. Therefore, a factor such as~\cite{krovi_improved_2023}
\begin{equation}
    g=\frac{\max_{t\in[0,T]}\|u(t)\|}{\|u(T)\|}
\end{equation}
or refinements theerof~\cite{jennings2024cost} will likely contribute a term exponential in $T$ to the complexity of measuring the final timestep. 
We emphasise that this is a consequence of the interplay between the quantum linear solver~\cite{harrow2009quantum}, the nonlinear nature of the Vlasov-Poisson system, and the physical limits on the collision frequency in~\eqref{nu ub}.
The latter necessitates the shift by the Maxwellian for convergence of the Carleman embedded Vlasov-Poisson system, which in turn seems to increase the complexity of extracting time-resolved quantities exponentially.

\textit{Beyond Fourier-Hermite expansions}:
We now clarify the extent to which our conclusions are particular to the choice of Fourier-Hermite expansions. We define the general basis function expansion
\begin{equation}\label{general basis expansion}
    \hat{f}(x,v,t)= \sum_{n,m}C_{n,m}(t)g_n(x)h_m(v),
\end{equation}
assuming periodicity of the spatial functions $g_n(0)=g_n(L)$ and orthonormality 
\begin{align}\label{general x basis}
    \langle g_p, g_n\rangle_x = \int_0^L g_n(x)g_p(x)^*dx = \delta_{n,p},\\ \label{general v basis}
    \langle h_r, h_m\rangle_v = \int_\mathbb{R} h_m(v)h_r(v)^*dx = \delta_{m,r}
\end{align}
for the respective inner products. 
For any such basis functions, meeting the convergence criterion in the lower bound~\eqref{shifted_converg_identity} gives
\begin{equation} \label{general lb}
   \nu > C\tau(N_v) 
\end{equation}
where the constant $C>0$ and increasing function $\tau(N_v)$ depend on the choice of both basis functions and initial conditions.
The derivation is shown in Appendix \ref{proof of nu lb}.

In particular, the Fourier-Hermite basis we used earlier gives $\tau(N_v)=\sqrt{N_v}$. Similarly, a central finite difference method approximating the derivative with $\partial_v f\approx (f_{i+1}-f_{i-1})/2\Delta v$ gives a linear growth rate $\tau(N_v)=N_v$.
This suggests that a judicious choice of basis functions can play an important role in determining the convergence of the Carleman-embedded Vlasov-Poisson system.
\textit{Conclusion}:
We have demonstrated that convergence can be guaranteed for solving the shifted Vlasov-Poisson system for certain initial conditions and physically relevant collision frequencies using Carleman embedding. This is an advance over an earlier work that required unphysically large collision frequencies~\cite{vaszary2024solving}.

Our guarantee holds for initial conditions close to the Maxwellian distribution. 
As they move away from the Maxwellian, the number of velocity modes kept in the expansion has to be reduced to ensure the collision frequency does not exceed the physical limit. We show that this is not an artefact of a specific choice basis functions, but true for a large class.
We also identity a possible tradeoff between the convergence and complexity of the resulting quantum algorithm on extracting  time-resolved outputs. This can be traced back to the competition between the nonlinearity and the physically limits of the collision frequency.

Future work could expand the set of convergence criteria available for the Carleman embedding method to regimes of stronger nonlinearity or lower collision frequencies. This method may then be able to capture richer nonlinear phenomena that plasma physics has to offer. 
Analytical bounds on the minimized $\nu_P$ in Figure \ref{fig:shifted comparison} would be valuable. Those in the limit of large $N_v$ would be most illuminating, clarifying for instance, the behaviour in Figure \ref{fig:shifted comparison} (b).
Finally, it would be useful to know how fundamental the tradeoff between the collision frequency strength, nonlinearity, and complexity of extracting time-resolved outputs is.

\textit{Note}: During the completion of this manuscript, a related work on solving the Vlasov-Poisson system for kinetic ions using Carleman embedding appeared~\cite{berntson2026end}. It established convergence for weak nonlinearities analytically and provides a quantum algorithm with superquadratic speedup to extract the phase-space and time-averaged kinetic energy of the ions.

\textit{Acknowledgements}:
We thank James Kermode for helpful discussions on the numerical minimization algorithm. We thank Tam\'{a}s Vaszary and Elena Fern\'{a}ndez-Victorio Castro for helpful discussion on the application of quantum algorithms to plasma physics. 
MC thanks the Hetsys CDT for support.
AD was supported, in part, by the Hub for Quantum Computing via Integrated and Interconnected Implementations (QCI3) (EP/Z53318X/1). TG was supported, in part, by the UK Programme of Laser inertial Fusion Technology for energy sponsored by DESNZ.
Computing facilities were provided by the Scientific Computing Research Technology Platform of the University of Warwick.

\bibliography{main.bib}

\appendix

\section{Plasma Physics Details}\label{plasma physics appendix}
The dimensionless quantities in Eq.~\eqref{vlasov} are as follows:
\begin{equation}
\begin{split}
    x \rightarrow \frac{x}{\lambda_D},\quad
    v &\rightarrow \frac{v}{v_{th}},\quad
    t \rightarrow \omega_pt,\\
    f \rightarrow \frac{v_{th}}{n_0}f,\quad
    E &\rightarrow \frac{\lambda_D e}{k_B T_e}E,\quad
    \nu \rightarrow \frac{\nu}{\omega_p}.
\end{split}
\end{equation}
Here the plasma frequency $\omega_p = (n_0e^2/m_e\varepsilon_0)^{1/2}$ is dependent on the number density $n_0$, electron charge and mass $e$ and $m_e$, and temperature $T_e$. The Debye length $\lambda_D = (\varepsilon_0 k_B T_e/n_0 e^2)^{1/2}$ is a function of Boltzman constant $k_B$. The plasma frequency, Debye length and thermal velocity $v_{th}=(k_B T_e/m_e)^{1/2}$ are related by the equality $v_{th}=\lambda_D \omega_p$. 

The electron-electron collision frequency is given by~\cite{huba1998nrl}:
\begin{equation}
    \nu_{ee} = \frac{e^4 n_0}{12 \pi^{3/2} \epsilon_0^2 m_e^{1/2} T_e^{3/2}} \ln \Lambda_c.
\end{equation}
Many variations of the Coulomb logarithm, $\ln \Lambda_c$ exist in the literature~\cite{boyd2003physics, huba1998nrl}, but in its simplest form, for a classical plasma (i.e., neglecting modifications based on the de Broglie wavelength) it is given by
\begin{equation}
    \ln \Lambda_c = \ln \left(\frac{b_\mathrm{max}}{b_\mathrm{min}}\right) = \ln \left(\frac{\lambda_D}{r_c}\right) = \ln\Lambda.
\end{equation}
Here $r_c$ is the classical radius of closest approach obtained by equating the electron kinetic energy with the electrostatic force, and $\Lambda$ is the plasma parameter~\cite{chen2015introduction}
\begin{equation}
    \Lambda = \frac{4 \pi \epsilon_0^{3/2} k_B^{3/2} T_e^{3/2}}{3 e^3 n_0^{1/2}}.
\end{equation}
The dimensionless electron-electron collision frequency is 
\begin{equation}
    \frac{\nu_{ee}}{\omega_p} = \frac{1}{\Lambda} \ln \Lambda \frac{1}{9 \pi^{1/2}} \lesssim 1.
\end{equation}

We have taken the simplest possible version of the Coulomb logarithm. We have also neglected electron-ion collisions. One solution to this is to use the collision fix~\cite{epperlein1991practical}, $\bar\tau' = \bar\tau \gamma(z)$, to combine electron-electron and electron-ion collisions into a single term, which results in a similar limit on the rescaled collision frequency.

\section{Fourier-Hermite Expansion}\label{fh derivation}
We substitute the expansion
\begin{equation}
    \hat{f}(x,v,t) = \sum_{n,m}C_{n,m}(t)e^{inx}\frac{He_{m}(v)f_\text{M}(v)}{\sqrt{m!}},
\end{equation}
into Eqs. \eqref{shifted vlasov} \& \eqref{shifted poisson} and use the orthogonality relationship of the Fourier-Hermite modes to get an ODE in the coefficients of the expansion. 

The orthogonality of the Fourier modes gives,
\begin{equation}
    \int_0^{2\pi}e^{inx}e^{ipx}dx=\delta_{n,p}.
\end{equation}
and similarly for the Hermite polynomials we have, 
\begin{equation}
    \int_\mathbb{R}He_m(v)He_r(v)f_\text{M}dv=m!\delta_{m,r}
\end{equation}
where $f_{\text{M}}$ gives the weight function $w(v)=e^{v^2/2}$ required for the orthogonality of the Hermite polynomials.

The time derivative of the distribution function becomes
\begin{equation}
    \frac{\partial f}{\partial t} = \sum_{n,m}\dot{C}_{n,m}(t)e^{inx}\frac{He_m(v)}{\sqrt{m!}}f_{\text{M}},
\end{equation}
which when multiplied by $e^{ipx}\frac{He_r(v)}{\sqrt{r!}}$ and integrating gives the projected time derivative $\dot{C}_{p,r}$.

The linear spatial advection operator expands to be,
\begin{equation}
    v\frac{\partial f}{\partial x} = \sum_{n,m}C_{n,m}ine^{inx}\frac{vHe_m(v)}{\sqrt{m!}}f_{\text{M}}.
\end{equation}
Using the recurrence relation $vHe_m(v)=He_{m+1}(v)+mHe_{m-1}(v)$, the projection of the spatial advection operator becomes,
\begin{equation}
    ip(\sqrt{r+1}C_{p,r+1} + \sqrt{r}C_{p,r-1}).
\end{equation}

The nonlinear velocity advection couples the Vlasov equation with the Poisson equation. Fourier expanding the electric field $E=\sum_{n'} \mathcal{E}_{n'}(t)e^{in'x}$, Poisson's equation becomes
\begin{equation}
    \sum_{s}is\mathcal{E}_se^{isx} = -\sum_n C_{n,0}e^{inx}
\end{equation}
which gives $\mathcal{E}_{s}=\frac{i}{s}C_{s,0}$ for $s\neq0$ (taking the value $0$ when $s=0$). Substituting this into the velocity advection term, we get
\begin{equation}
    E\frac{\partial f}{\partial v} = \sum_{n,m,s}C_{s,0}C_{n,m}e^{i(n+s)x}\frac{mHe_{m}}{\sqrt{m!}}f_{\text{M}}
\end{equation}
where we have used that $\partial_vHe_m(v)=mHe_{m-1}(v)$. Once again, using the above orthogonality relationship gives the projected convolution term,
\begin{equation}
    \sum_{s\neq0}\frac{i\sqrt{r}}{s}C_{s,0}C_{p-s,r-1}.
\end{equation}

The Maxwellain advection term $Evf_\text{M}$ is linear in the distribution function and expands to become,
\begin{equation}
    Evf_\text{M} = \sum_{s\neq 0 } \frac{i}{s}C_{s,0}e^{isx}vf_{\text{M}}.
\end{equation}
Noting that $v=He_1(v)$, the projection is then
\begin{equation}
    \frac{i}{p}C_{p,0}\delta_{r,1}.
\end{equation}

Finally, the collision operator is trivially projected into the Fourier-Hermite space, resulting in a contribution,
\begin{equation}
    -\nu C_{p,r},
\end{equation}
to the projected equation.

\section{Mapping from Fourier-Hermite to general quadratic nonlinear form}\label{mappings}
To use the convergence criteria provided by \cite{jennings2025quantum}, we need to map the linear and nonlinear terms of the Fourier-Hermite formulation to the matrices $F_1$ and $F_2$ respectively.

\subsection{Linear terms}
There are three contributions to the linear $F_1$ matrix: the linear spatial advection, the collision operator and Maxwellian advection term. When projected in the Fourier-Hermite basis, the spatial advection operator adds the terms $-ik_p(\sqrt{r}C_{p,r-1}+\sqrt{r+1}C_{p,r+1})$ to Eq. \eqref{shift. VP mode eq}. This can be mapped to the $F_1$ matrix by the term $-iS=-iG\otimes K$, where
\begin{equation}
    G = \begin{pmatrix}
0 & 1 &  &  &  \\
1 & 0 & \sqrt{2} &  &  \\
 & \sqrt{2} & \ddots & \ddots &  \\
 &  & \ddots & 0 & \sqrt{N_v} \\
  &  &  & \sqrt{N_v} & 0
\end{pmatrix}
\end{equation}
and $K=\text{diag}(-N_x/2, -N_x/2+1,\dots, N_x/2)$. The contribution from the collision operator is $-\nu I$. Finally, the Maxwellian advection contributes the matrix, 
\begin{equation}
    -iY = -i\begin{pmatrix}
0      &        &        &        &   \\
D      & 0      &        &        &   \\
       & \ddots & \ddots &        &   \\
       &        & 0      & 0      &   \\
       &        &        & 0      & 0
\end{pmatrix},
\end{equation}
where $D = \text{diag}\!\left(\frac{-1}{N_x}, \frac{-1}{N_x-1}, \ldots, 1, 0, 1, \ldots, \frac{1}{N_x}\right)$. Hence, the linear matrix is the summation of these matrices $F_1 = -\nu I -i(S+Y)$.

\subsection{Nonlinear terms}
The nonlinear contributions come from the coupling of the velocity advection to the electric field evaluated using Poisson's equation. The contributions in Eq. \eqref{shift. VP mode eq} are of the form $\sum_{s\neq0}\frac{i\sqrt{r}}{k_s}\hat{C}_{s,0}\hat{C}_{p-s,r-1}$. The $F_2$ is a rectangular matrix, mapping $u^{\otimes2}\mapsto u$, so we introduce the indexing function $(x, y)= y\times(N_x+1) + x$. Defining the two indices $j_1=(s,0)$ and $j_2=(p-s, r-1)$ and letting $N=(N_x+1)(N_v+1)$ being the total system size, we write the $F_2$ matrix elementwise as 

\begin{equation}
    [F_2]_{i,\,j_1N+j_2} = \begin{cases} 
-\dfrac{i\sqrt{r}}{s} & \text{if } i=(p,r),\; j_1=(s,0),\; 
    \\&i_2=(p-s,r-1),\; \\&s\neq 0,\; r\geq 1, \\[10pt] 
0 & \text{otherwise.} 
\end{cases}
\end{equation}

\section{Computational Details} \label{computation details}
To minimize the bounds over the space of Hermitian positive definite matrices, we employ Riemannian optimization methods. The Python package \texttt{Pymanopt} \cite{manopt} is used to find local minima of the lower bounds from multiple randomly sampled matrices, using a conjugate gradient optimizer to find these minima. We use $N_{start}=100$ random samples to generate the local minima, and take the lowest of these values as the numerically minimised lower bound. A gradient descent minimizer was used to find the local minima of the cost function. 

The matrices $F_1$ and $F_2$ were populated according to the desciptions in Appendix \ref{mappings}. The spectral and $l_2$ norms were evaluated numerically using \texttt{numpy} and the machine learning package \texttt{JAX} \cite{jax2018github} was used to calculate gradients of the cost function.

To calculate the $P$ log-norms, we reconstruct it as the following eigenvalue problem,
\begin{align*}
    \mu_P(F_1) &= \max_{x\neq0}\Re\left\{\frac{\langle F_1x,x\rangle _P}{\langle x,x\rangle _P}\right\}\\
    &=\max_{x\neq0}\frac{x^\dagger(PF_1+F_1^\dagger P)x}{2x^\dagger Px}\\
    &=\max_{y\neq0}\frac{y^\dagger P^{-1/2}(PF_1+F_1^\dagger P)P^{-1/2}y}{2y^\dagger y}\\
    &=\max_{y\neq0}\frac{y^\dagger(P^{1/2}F_1 P^{-1/2}+P^{-1/2}F_1^\dagger P^{1/2})y}{2y^\dagger y}\\
    &=\lambda_{\text{max}}\left(\frac{P^{1/2}F_1 P^{-1/2}+P^{-1/2}F_1^\dagger P^{1/2}}{2} \right)
\end{align*}
where we have introduced $y=P^{1/2}x$.

\section{Direct Application}\label{direct}
Here we show the results of the same minimization algorithm used for the shifted set of Eqs. \eqref{shift. VP mode eq}. We find that although the minimization algorithm has a much greater effect, none of the bounds sit below the physical limit of inequality \eqref{nu ub}.

\begin{figure*}
    \centering
    \includegraphics[width=\linewidth]{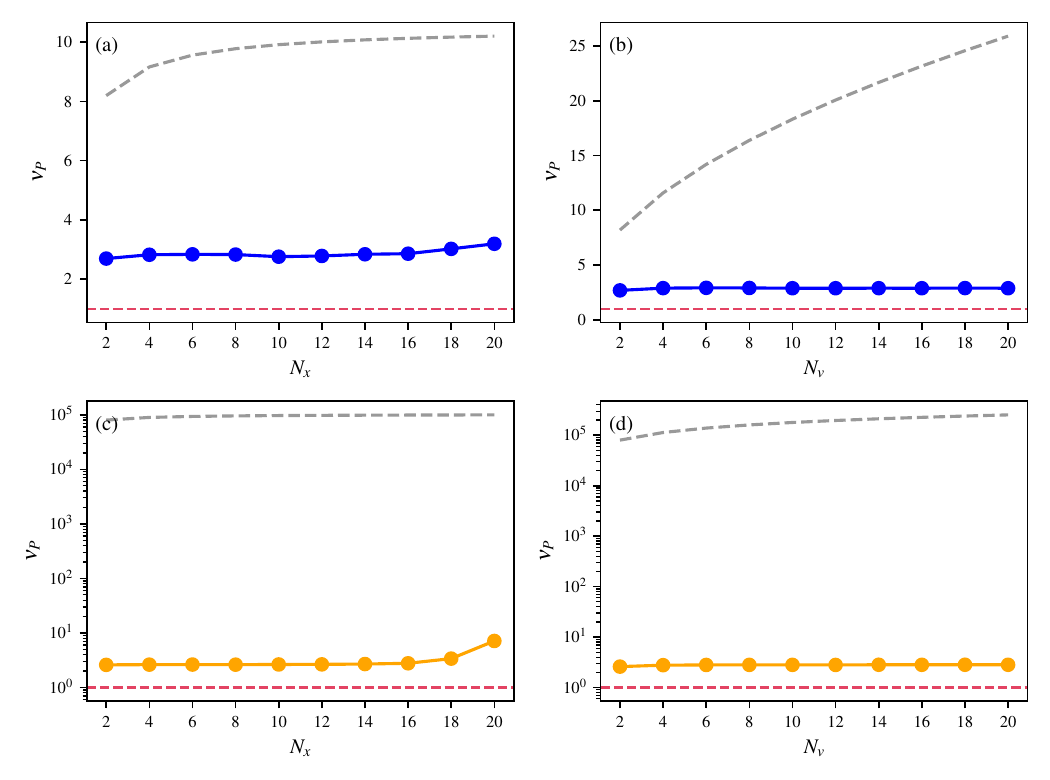}
    \caption{Numerically obtained lower bounds of collision frequency in the unshifted regime (blue \& yellow), the fixed $P=I$ bounds (grey) and physical upper limit $\nu<1$ (red). (a) Two stream instability initial condition with fixed $N_v=2$. (b) Two stream instability initial condition with fixed $N_x=2$. (c) Perturbed Maxwellian initial condition with fixed $N_v=2$. (d) Perturbed Maxwellian initial condition with fixed $N_x=2$.}\label{fig:unshifted_comparison}
\end{figure*}

The Fourier-Hermite expansion applied directly to Eqs.~\eqref{vlasov} \& \eqref{poisson} is given by 
\begin{equation}\label{VP mode eq}
\begin{split}
    &\dot{C}_{p,r}+ip(\sqrt{r}C_{p,r-1}+\sqrt{r+1}C_{p,r+1})\\&+\sum_{s\neq0}\frac{i\sqrt{r}}{s}C_{s,0}C_{p-s,r-1} = -\nu(C_{p,r}-\delta_{p,0}\delta_{r,0}).
\end{split}
\end{equation}
The equivalent lower bound for the unshifted Fourier-Hermite system is
\begin{equation} \label{direct cf P lb}
    \nu > \frac{\|u_0\|_P\left(\mu_P(-iS) + \|F_2\|_P\|u_0\|_P\right)}{\|u_0\|_P-\|P_M\|_P} = \nu_P,
\end{equation}
where $P_M$ is the projection of the Maxwellian into the Fourier-Hermite basis.

\subsection{Numerically Minimised $P$}
Using the same minimization algorithm as for the shifted equations, we obtain the minimized bounds in Figure \ref{fig:unshifted_comparison}.

For the two stream instability initial condition, the numerical minimization provides bounds that are reduced for all mode numbers tested. This improvement is more pronounced for the perturbed Maxwellian initial condition, saving orders of magnitude on the required bound for the collision frequency. However, in both cases the numerical bounds sit above the physical $\nu<1$ limit.

\section{Norm Evaluations}\label{norm evaluations}
The log-norm $\mu(-i(S+Y))$ can be explicitly calculated,
\begin{align*}
    \mu(-i(S+Y)) &= \mu(-iY)= \frac{1}{2}
\end{align*} 
where we have used the symmetry of $S$ to eliminate it from this term.

The spectral norm of $F_2$ can be calculated explicitly by first taking its Hermitian adjoint $F_2^\dagger$. This is advantageous as $F_2 F_2^\dagger$ is diagonal, whereas $F_2^\dagger F_2$ is not. Therefore,
\begin{align*}
    \|F_2\|_2= \|F_2^\dagger\|_2&=\sqrt{\lambda_{\max}(F_2F_2^\dagger)} =\sqrt{\max_{j}[F_2F_2^\dagger]_{j,j}}\\
    &=\sqrt{\max_{p,r}r\sum_{\text{valid } s} \frac{1}{s^2}}\\ &=\sqrt{N_v}\cdot\max_{p}\sqrt{\sum_{\text{valid } s} \frac{1}{s^2}}=\sqrt{2N_v\sum_{s=1}^{N_x/2}\frac{1}{s^2}}
\end{align*}
where valid $s$ is the set satisfying $\{s\neq0, |s| \leq N_x/2, |p-s| \leq N_x/2\}$. The summation over the spatial modes 
is upper bounded by $\pi^2/3$, so asymptotically the spectral norm $\|F_2\|_2 = \mathcal{O}(\sqrt{N_v})$. 

Finally, the $\ell^2$ norm $\|u_0\|_2$ depends on the choice of the initial condition. 
Taking $u_0 = f_{\text{pert}}$, we get
\begin{equation}
    \|u_0\|_2 = \frac{\alpha}{\sqrt{2}}.
\end{equation}
Alternatively, setting $u_0 = f_{\text{ts}}$, the norm becomes
\begin{equation}
    \|u_0\|_2=\sqrt{2+\frac{3\alpha^2}{2}}.
\end{equation}

The grey dashed lines in Figure \ref{fig:shifted comparison} are then given by the analytical expression 
\begin{equation}
    \nu_I=\frac{1}{2}+\frac{\alpha}{2}\sqrt{N_v\sum_{s=1}^{N_x/2}\frac{1}{s^2}}
\end{equation}
for the perturbed Maxwellian and
\begin{equation}
\nu_I=\frac{1}{2}+\sqrt{\left(2+\frac{3\alpha^2}{2}\right)N_v\sum_{s=1}^{N_x/2}\frac{1}{s^2}}
\end{equation}
for the two stream instability.

\section{Basis Agnostic $\nu$ Lower Bound}\label{proof of nu lb}

Using the general basis defined in \eqref{general basis expansion}, we can write the Vlasov-Poisson as 
\begin{equation}
    \frac{\partial u}{dt}= \underbrace{(-\nu I + \mathcal{S} + \mathcal{T})}_{F_1} u+F_2(u\otimes u),
\end{equation}
where $\mathcal{S}$ is the general $x$ advection operator and $\mathcal{T}$ is the contribution from the advection of the Maxwellian term. 

\subsection{Log-Norm $\mu(F_1)$}
We can analyze the three contributions to the log-norm individually. The collision operator adds a trivial shift of $-\nu$ to the log-norm as all modes are equally damped. 

The spatial advection operator can be written in terms of its two components $\mathcal{S}=\mathcal{V}\otimes\mathcal{X}$, where
\begin{align}
    &[\mathcal{X}]_{p,n}=\langle g_p, g_n' \rangle_x, \\
    &[\mathcal{V}]_{r,m}=\langle h_r, vh_m \rangle_v.
\end{align}
Due to the periodicty assumption of the spatial basis functions, we have
\begin{align*}
    [\mathcal{X}]_{p,n}  &= \int_0^L g_n'(x)g_p(x)^*dx\\
    &=[g_n(x)g_p(x)^*]_0^L - \int_0^L g_n(x)g'_p(x)^*dx\\
    &=- \left( \int_0^L g'_p(x)g_n(x)^*dx\right)^*\\
    &=-[\mathcal{X}]_{n,p}^*
    = -[\mathcal{X}^\dagger]_{p,n}
\end{align*}
hence $\mathcal{X}^\dagger=-\mathcal{X}$ and $\mathcal{X}$ is a skew-Hermitian operator. Inspecting $\mathcal{V}$, we see that
\begin{align*}
    [\mathcal{V}]_{r,m} &= \int_\mathbb{R} vh_m(v)h_r(v)^*dv \\
    &= \left(\int_\mathbb{R} vh_r(v)h_m(v)^*dv \right)^*\\
    &= [\mathcal{V}]_{m,r}^*
    =[\mathcal{V}^\dagger]_{r,m}
\end{align*}
therefore $\mathcal{V}^\dagger = \mathcal{V}$ is a Hermitian operator. Then $\mathcal{S}=\mathcal{V}\otimes \mathcal{X}$ is a Skew-Hermitian operator. 

The final contribution to the log-norm is from the Maxwellian advection term. In the general expansion of the basis, it can be expressed as $\mathcal{T} = (\mathcal{Z}\mathcal{W}^\top)\otimes \Gamma$. The spatial contribution $\Gamma$ originates from the $x$ integration of Poisson's equation and has the entries 
\begin{align}\label{gamma inner prod}
    [\Gamma]_{p,n}&=\langle g_p, \mathcal{N}_n(x)\rangle_x\\
    & =\int_0^L\int_0^x g_n(x)dx g_p(x)^*dx
\end{align}
The velocity part $\mathcal{Z}\mathcal{W}^\top$ can be expressed as an outer product between two vectors $\mathcal{Z}$ and $\mathcal{W}$, with entries 
\begin{align}
    &[\mathcal{Z}]_{r} = \langle h_r, \partial_v f_\text{M}\rangle_v,\\
    &[\mathcal{W}]_{m} = \langle 1, h_m\rangle_v, \label{1 inner prod}
\end{align}
so $\text{rank}(\mathcal{Z}\mathcal{W}^\top)=1$. Then, the rank of $\mathcal{T}+\mathcal{T}^\dagger$ is bounded by
\begin{align*}
    \text{rank}(\mathcal{T}+\mathcal{T}^\dagger)&\leq 2\text{rank}(\mathcal{T})\\
    &\leq 2\text{rank}((\mathcal{Z}\mathcal{W}^\top)\otimes\Gamma)\\
    &=2\text{rank}(\mathcal{Z}\mathcal{W}^\top)\text{rank}(\Gamma)\\
    &=2\text{rank}(\Gamma)\\
    &\leq 2(N_x+1).
\end{align*}
Hence if we take $N_v > 1$, 
\begin{align*}
    \dim(\mathcal{T}+\mathcal{T}^\dagger) &= (N_x+1)(N_v+1)\\
    &>2(N_x+1)\\
    &\geq\text{rank}(\mathcal{T}+\mathcal{T}^\dagger)
\end{align*}
and $\mathcal{T}+\mathcal{T}^\dagger$ must be singular. Therefore $\lambda_{\max}(\mathcal{T}+\mathcal{T}^\dagger)\geq 0$.

The log-norm of the linear terms can then be shown to be 
\begin{align*}
    \mu(F_1) &= \mu(-\nu I +\mathcal{S}+\mathcal{T})\\
    &=\frac{1}{2}\lambda_{\max}(-\nu I +\mathcal{S}+\mathcal{T}+(-\nu I +\mathcal{S}+\mathcal{T})^\dagger)\\
    &=-\nu + \frac{1}{2}\lambda_{\max}(\mathcal{S}+\mathcal{S}^\dagger + \mathcal{T} + \mathcal{T}^\dagger)\\
    &=-\nu + \frac{1}{2}\lambda_{\max}(\mathcal{T} + \mathcal{T}^\dagger) 
    \geq -\nu.
\end{align*}
We can then use this to lower bound the $R_I$ values,
\begin{equation}
    \frac{\|F_2\|_2 \|u_0\|}{\nu} \leq R_I <1,
\end{equation}
so for the convergence criteria to be satisfied, $\nu > \|F_2\|_2\|u_0\|_2$.

\subsection{Initial State Norm}
Projecting the initial condition into the general basis, the vector $u_0$ is given by the entries
\begin{equation}
    [u_0]_{(n,m)} = \int_0^L\int_\mathbb{R}f(x,v,0)g_n(x)^*h_m(v)^*dvdx.
\end{equation}
For the dynamics to be non-trivial, at least one of the entries of $u_0$ has to be nonzero. Taking $C_1>0$ to be the absolute value of this minimum nonzero entry, we have
\begin{equation}
    \nu > C_1\|F_2\|_2.
\end{equation}

\subsection{Nonlinear Norm}
The $F_2$ matrix containing the nonlinear contributions to the ODEs can be written (up to column reordering) as 
\begin{equation}
    F_2 = \mathcal{D}\otimes\mathcal{W}^\top\otimes\Gamma,
\end{equation}
where $[\mathcal{D}]_{r,m}=\langle h_r, h_m'\rangle$ is the projection of the velocity derivative into the basis, $\mathcal{W}$ is defined in \eqref{1 inner prod} and $\Gamma$ is defined in \eqref{gamma inner prod}. The norm of $F_2$ can be factored into its individual components $\|F_2\|_2 = \|\mathcal{D}\|_2\|\mathcal{W}\|_2\|\Gamma\|_2$. Setting $\tilde{w}$ and $\tilde{\gamma}$ to be nonzero elements of $\mathcal{W}$ and $\Gamma$ with fixed mode numbers $n,n',p$ and $m'$, and defining $C_2 = \tilde{w}\tilde{\gamma}$, it is clear that 
\begin{equation}
    \|F_2\|_2 \geq C_2\|\mathcal{D}\|_2,
\end{equation}
where $C_2$ depends only on the choice of basis and is independent of $N_x$ and $N_v$. Since $\mathcal{D}$ has elements $[\mathcal{D}]_{r,m}=\langle h_r, h_m'\rangle$, its spectral norm is lower bounded by the maximum absolute value of its elements,
\begin{equation}
    \|\mathcal{D}\|_2\geq\max_{i,j}|\langle h_{i},h'_{j}\rangle_v|=\tau(N_v),
\end{equation}
which we show is an increasing function of $N_v$ for a number of basis function choices.

Therefore,
\begin{align*}
    \nu &> C_1\|F_2\|_2
    \geq C_1C_2\|\mathcal{D}\|_2\\
    &\geq C_1C_2 \max_{i,j}|\langle h_i, h'_j \rangle_v|
    =C_1C_2\tau(N_v).
\end{align*}
Hence letting $C=C_1C_2$ gives the result found in inequality \eqref{general lb}.

\subsubsection{Hermite Polynomials}
For the Hermite polynomials, the symmetric expansion becomes $h_m(v)=He_m(v)f_\text{M}^{1/2}$, the velocity derivative is $h'_m(v)=\frac{1}{2}(\sqrt{m}h_{m-1}(v)-\sqrt{m+1}h_{m-1}(v))$. Then, the maximum element of $\mathcal{D}$ is 
\begin{equation}
    \max_{i,j}|\langle h_{i},h'_{j}\rangle_v| = \sqrt{N_v},
\end{equation}
which is the same result as in Appendix~\ref{norm evaluations} but for the symmetrized basis function.

\subsubsection{Finite Difference Discretisation}
Approximating the velocity derivative with the central difference $\partial_v f \approx\frac{f(x,v_{i+1}t) - f(x,v_{i-1}t)}{2\Delta v}$ \cite{vaszary2024solving}. The norm of $\mathcal{D}$ is then $\|\mathcal{D}\|_2 = \frac{1}{\Delta v} = \frac{N_v}{L_v}$ where $L_v$ is the length of the velocity grid. Therefore, the rate of growth is linear in the number of velocity points $\tau(N_v) = N_v$. For convenience we have also assumed that the grid points have been chosen to land on the same points in velocity space as the number of points is increased. This is without loss of generality.

\end{document}